\documentclass{aa}

\usepackage{txfonts}

\usepackage{graphicx}

\usepackage{array}

\usepackage{epsfig}

\usepackage{natbib}
\bibpunct{(}{)}{;}{a}{}{,} 

\newcommand{\teff}{$T_\mathrm{eff}$}
\newcommand{\logg}{$\log g$}

\newcommand{\prot}{$P_\mathrm{rot}$}

\newcommand{\rwd}{$R_\mathrm{WD}$}

\newcommand{\bdip}{$B^\mathrm{d}_\mathrm{pol}$} 
\newcommand{\bqua}{$B^\mathrm{q}_\mathrm{pol}$}
\newcommand{\boct}{$B^\mathrm{o}_\mathrm{pol}$}
\newcommand{\thed}{$\Theta^\mathrm{d}$}
\newcommand{\phid}{$\Phi^\mathrm{d}$}
\newcommand{\theq}{$\Theta^\mathrm{q}$}
\newcommand{\phiq}{$\Phi^\mathrm{q}$}
\newcommand{\theo}{$\Theta^\mathrm{o}$}
\newcommand{\phio}{$\Phi^\mathrm{o}$}

\newcommand{\xoff}{${x'}_\mathrm{off}$}
\newcommand{\yoff}{${y'}_\mathrm{off}$}
\newcommand{\zoff}{${z'}_\mathrm{off}$}

\newcommand{\chisq}{$\chi^2$}
\newcommand{\chisqred}{$\chi^2_\mathrm{red}$}

\newcommand{\bpd}{\mbox{$B$--$\psi$ diagram}}

\newcommand{\lbc}{\mbox{$\lambda$--$B$ curves}}

\newcommand{\glm}{$g_l^m$}
\newcommand{\hlm}{$h_l^m$}

\newcommand{\ohe}{\mbox{\object{HE\,1045$-$0908}}}

\begin{document}

\title{Zeeman tomography of magnetic white dwarfs}
\subtitle{II.\ The quadrupole-dominated magnetic field of \ohe\thanks{
Based on observations collected at the European Southern Observatory,
Paranal, Chile, under programme IDs \mbox{63.P-0003(A)} 
and \mbox{64.P-0150(C)}.}}

   \author{F.~Euchner\inst{1} \and
        K.~Reinsch\inst{1} \and
	S.~Jordan\inst{2} \and
	K.~Beuermann\inst{1} \and
        B.\,T.~G\"ansicke\inst{3}}

        \offprints{F.~Euchner, \email{feuchner@astro.physik.uni-goettingen.de}}
   
        \institute{Institut f\"ur Astrophysik, Universit\"at G\"ottingen,
        \mbox{Friedrich-Hund-Platz~1}, \mbox{D-37077~G\"ottingen}, Germany 
	\and 
	Astronomisches Rechen-Institut am ZAH, M\"onchhofstr.~12--14, 
	D-69120~Heidelberg, Germany 
	\and 
	Department of Physics, University of Warwick, Coventry~CV4~7AL, UK}
             
\date{Received 11~March~2005 / Accepted 20~July~2005}

\abstract{We report 
time-resolved optical flux and
circular polarization spectroscopy of the magnetic DA white dwarf
\ohe\ obtained with FORS1 at the ESO VLT. 
Considering published results, we estimate
a likely rotational period of \mbox{\prot\ $\simeq 2.7$\,h}, but 
cannot exclude values as high as about 9\,h.
Our detailed Zeeman tomographic analysis reveals a field structure
which is dominated by a quadrupole and contains additional dipole and
octupole contributions, 
and which does not depend strongly on the assumed value 
of the period.
A good fit 
to the Zeeman flux and polarization spectra is obtained
if all field components are centred and inclinations of 
their magnetic axes
with respect to each other are allowed for. The fit can be slightly
improved if an offset from the centre of the star is included. The
prevailing surface field strength is 16\,MG, but values between 10 and
\mbox{$\sim$\,75\,MG} do occur. We derive an effective photospheric
temperature of \ohe\ of \mbox{\teff\ = 10\,000 $\pm$ 1000\,K}. The
tomographic code makes use of an extensive database of pre-computed
Zeeman spectra (Paper~I).
 }
   
\titlerunning{Zeeman tomography of magnetic white dwarfs.\ II.\ \ohe}
\authorrunning{F.~Euchner et al.}
\maketitle


\section{Introduction}

Until a few years ago, magnetism among white dwarfs had been considered
a rare phenomenon. 
A fraction of \mbox{$\sim$\,5\,\%}\ of all known white dwarfs had been
confirmed to be magnetic, with field
strengths covering the range from
\mbox{$\sim$\,30\,kG--1000\,MG}\footnote{1~MG = $10^6$~Gauss = 100~Tesla}
\citep{wickramasinghe+ferrario00-1}.
Presently, the low-field tail of the known field strength distribution
is established by four objects in the kilogauss range (\mbox{$B \simeq
2$--4\,kG}), which is the current detection limit for 8-m class
telescopes \citep{fabrika+valyavin99-1, aznarcuadradoetal04-1}.
Recent studies suggest a much higher fractional incidence of magnetic
white dwarfs (MWDs) of at least \mbox{10\,\%}\ for objects with
surface fields exceeding 2\,MG, and probably even more if low-field
objects are included (\citeauthor{liebertetal03-1}
\citeyear{liebertetal03-1}; \citeauthor{schmidtetal03-1}
\citeyear{schmidtetal03-1}; and references therein).
While a high incidence is mainly found for cool, old white dwarfs, it
is interesting to note that a high incidence of (weak) magnetic fields
has also been detected in central stars of planetary nebulae, which
are the direct progenitors of white dwarfs \citep{jordanetal05-1}, and
in subdwarf B and O~stars \citep{otooleetal05-2}.
There is strong evidence that the high-field magnetic white dwarfs
have evolved from main-sequence Ap and Bp stars. Low- and
intermediate-field objects are thought to originate either from late
A~stars that fall just above the mass limit below which fossil fields
are destroyed in the pre-main-sequence phase \citep{toutetal04-1}, or
from main sequence stars of still later spectral type.

Since there is no known mechanism to generate very strong magnetic
fields in white dwarf interiors, the fields are believed to be fossil
remnants of previous evolutionary stages \citep{braithwaite+spruit04-1}.
Modelling of the field evolution 
showed that the characteristic time for Ohmic decay of the lowest
poloidal multipole components is long compared with the white dwarf
evolutionary timescale \citep{wendelletal87-1, cumming02-1}.
Higher-order modes do not necessarily decay faster, however,
since they may be enhanced by
nonlinear coupling by the Hall effect if internal toroidal fields are
present \citep{muslimovetal95-1}. This is consistent with the finding
of significant deviations from pure dipole configurations
(\citeauthor{burleighetal99-1} \citeyear{burleighetal99-1};
\citeauthor{maxtedetal00-1} \citeyear{maxtedetal00-1};
\citeauthor{reimersetal04-1} \citeyear{reimersetal04-1};
\citeauthor{euchneretal05-1} \citeyear{euchneretal05-1}).

\begin{figure*}[t]
\includegraphics[bb=18 34 564 742,width=8.8cm,clip]{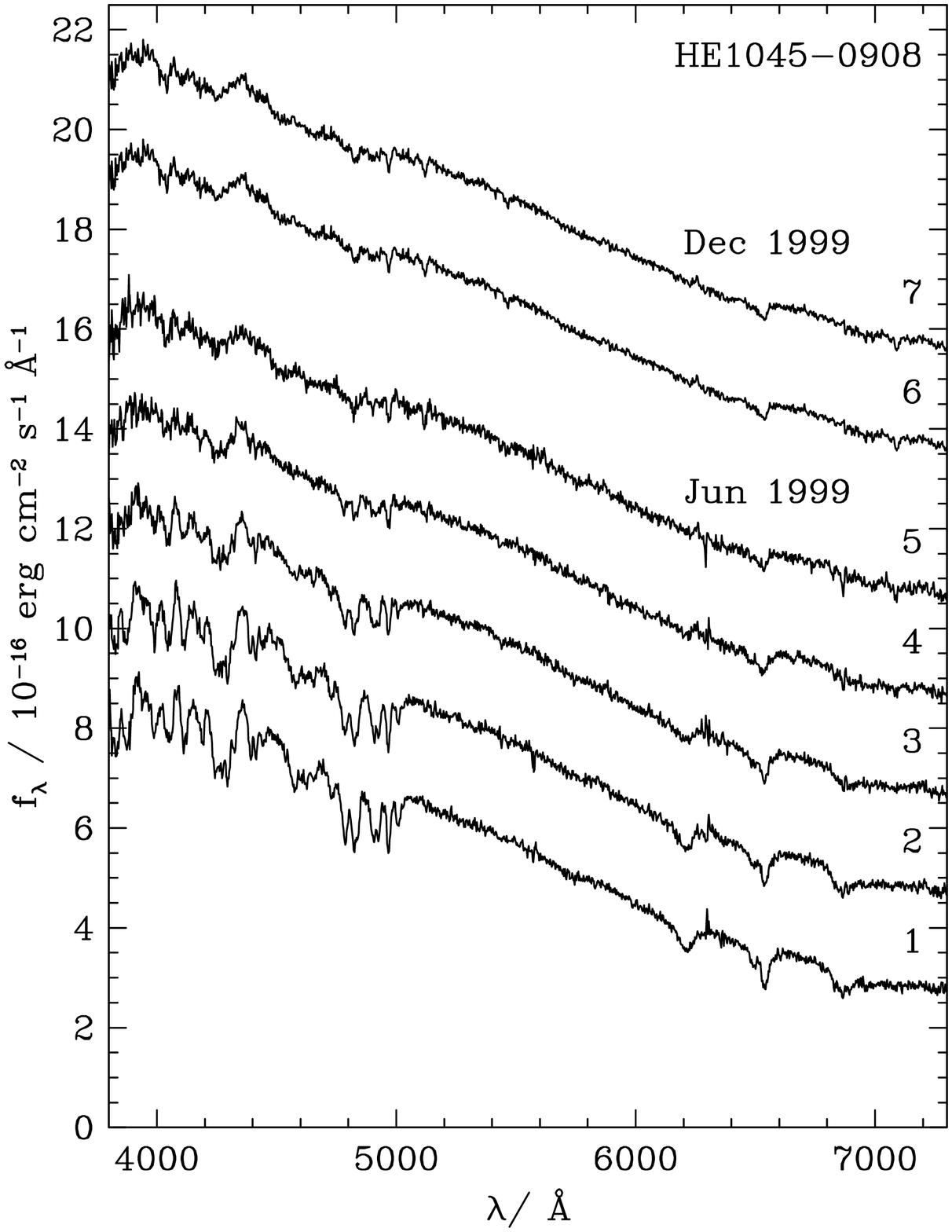}
\hfill
\includegraphics[bb=18 34 564 742,width=8.8cm,clip]{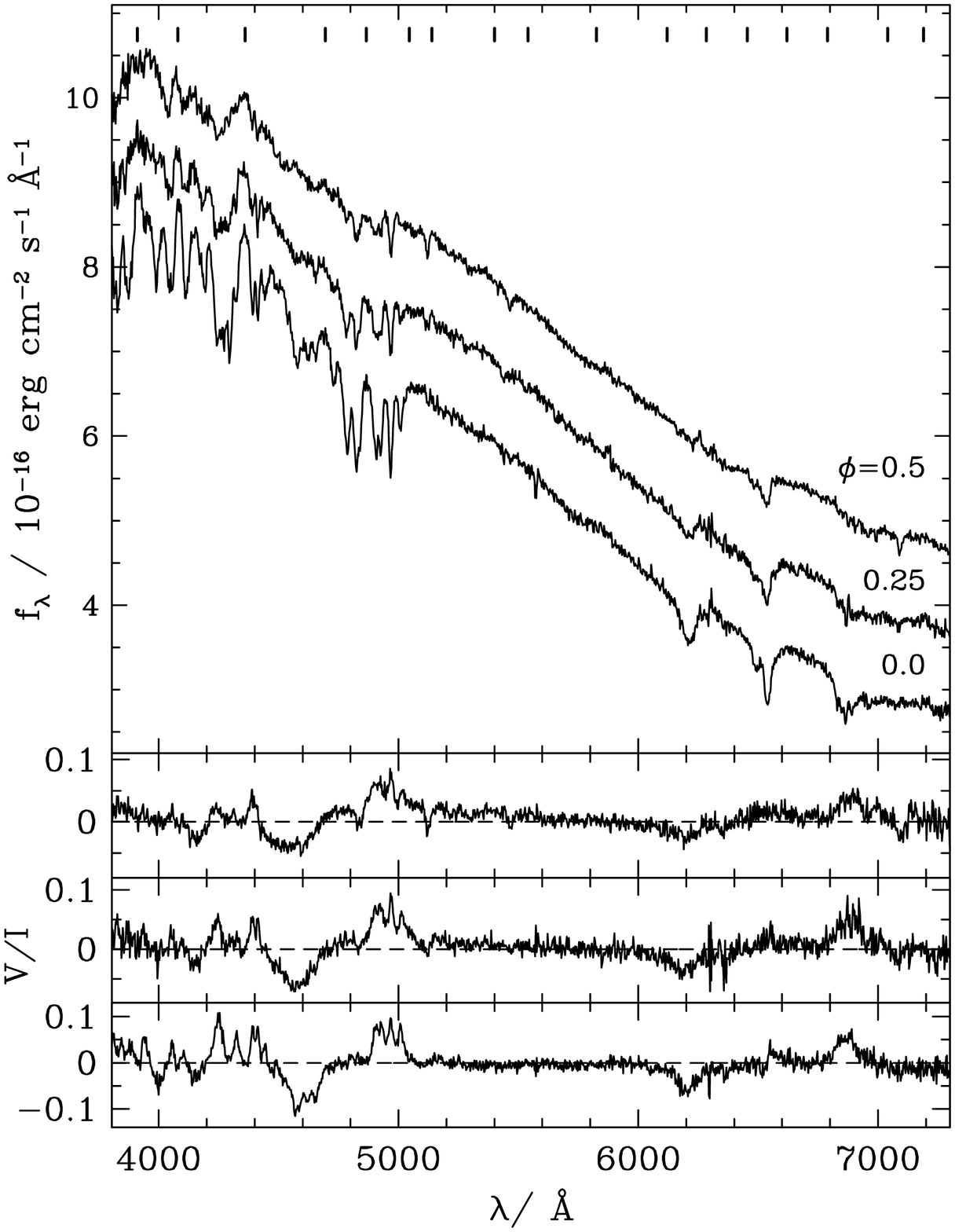}
\caption{\textit{Left panel:} Flux spectra of \ohe\ from June~1999
(1--5) and December~1999 (6--7). Spectra~2--7 have been shifted
upwards by two flux units each with an additional unit between
spectra~5 and 6.
\textit{Right panel:} Combined flux and circular
polarization spectra of \ohe\ from the June/December~1999 observations,
covering approximately one-half of the rotation cycle. These spectra,
which have been collected into three phase bins, have been used as input
spectra for the Zeeman tomographic procedure. For clarity, the
uppermost two curves have been shifted upwards by one and two flux units,
respectively. 
The quoted phases refer to case~(i) with \mbox{\prot\ = 2.7\,h}.}
\label{fig:he1045-raw-obs}
\end{figure*}

The present Zeeman tomographic analysis of phase-resolved
circular spectropolarimetry of the white dwarf \ohe\ provides further
evidence for strongly non-dipolar fields.
In the first paper of this series, we have demonstrated the
ability of our code
to derive the field configuration of rotating MWDs using
phase-resolved flux and circular polarization spectra
\citep[][ henceforth referred to as Paper~I]{euchneretal02-1}. In this
and follow-up papers, we apply our code to individual objects.  

\ohe\ was discovered in the Hamburg/ESO objective prism survey for
bright quasars \citep{wisotzkietal96-1}. Subsequent optical
spectroscopy at the ESO 3.6-m telescope revealed a rich spectrum of
Zeeman-split Balmer absorption lines and confirmed the object as a
magnetic DA white dwarf \citep{reimersetal94-1}.  By fitting
theoretical Zeeman spectra for a centred dipole with a
\mbox{trial-and-error} method, the best match was found by these
authors for \mbox{\teff\ = 9200\,K}, \mbox{\bdip\ = 31\,MG}, and a
nearly equator-on view.
\citet{schmidtetal01-1} subsequently obtained a sequence of five
flux and circular polarization spectra of \ohe\ over a duration of
1\,h.
The shape of the flux spectra in their observation sequence changes
monotonically from almost vanishing to strong Zeeman features, whereas
the variation in circular polarization is less pronounced.
They estimated that the 1-h interval represented either
one-quarter or one-half of a complete rotation cycle with a
probable rotational period of \mbox{\prot\ $\simeq 2$--4\,h}.


\section{Observations}
\label{sec:observations}

\begin{table}[b]
\caption{Dates and times for the spectropolarimetric observations of
\ohe\ obtained at the ESO VLT ($t_\mathrm{exp}$: exposure time, n:
number of exposures).}
\label{tab:obs-log}
\centering
\begin{tabular}{lllll} 
\hline \hline \noalign{\smallskip}

Object & Date & UT & $t_\mathrm{exp}$ (min) & n\\ \noalign{\smallskip} \hline
\noalign{\smallskip}
\ohe\	& 1999/06/09 & 22:55--00:21 & 20 & 4  \\
        & 1999/06/10 & 00:23--00:37 & 14 & 1  \\
	& 1999/12/06 & 08:27--08:44 & 8  & 2  \\
\noalign{\smallskip} \hline
\end{tabular}
\end{table}

We obtained rotational-phase resolved circular spectropolarimetry for
the magnetic DA white dwarf \ohe\ with FORS1 at the ESO VLT UT1/Antu
in June and December~1999.  The dates and times of the observations as
well as the number of exposures and exposure times are given in
Table~\ref{tab:obs-log}.
The spectrograph was equipped with a thinned, anti-reflection coated
2048$\times$2048-pixel Tektronix TK-2048EB4-1 CCD detector. For all
observations, the GRIS\_300V+10 grism with order separation filter
GG~375 covering the wavelength range \mbox{$\sim$\,3850--7500\,\AA}\
was used with a slit width of 1\arcsec\ yielding a FWHM spectral
resolution of \mbox{13\,\AA}\ at \mbox{5500\,\AA}.
We were able to reach a signal-to-noise ratio \mbox{$S/N \simeq 100$}
per resolution bin for the individual flux spectra.
The instrument was operated in spectropolarimetric (PMOS) mode. The
polarization optics consists of a Wollaston prism for beam separation
and two superachromatic phase retarder plate mosaics. Since both
plates cannot be used simultaneously, only the circular polarization
has been recorded using the quarter wave plate.
Spectra of
the target star and comparison stars in the field have been obtained
simultaneously by using the multi-object spectroscopy mode of
FORS1. This allows us to derive individual correction functions for the
atmospheric absorption losses in the target spectra and to check for
remnant instrumental polarization.


\subsection{Data reduction}

The observational data have been reduced according to standard
procedures (bias, flat field, night sky subtraction, wavelength
calibration, atmospheric extinction, flux calibration) using the
context MOS of the ESO MIDAS package.
In order to eliminate observational biases caused by Stokes parameter
crosstalk, the wavelength-dependent degree of circular polarization
$V/I$ has been computed from two consecutive exposures recorded with
the quarter wave retarder plate rotated by $\pm$45\degr\ according to
\begin{equation}
 \frac{V}{I} = \frac{1}{2} \left[ 
\left( \frac{f^\mathrm{o} - f^\mathrm{e}}{f^\mathrm{o} + 
f^\mathrm{e}} \right)_{\theta = 45^\circ} - 
\left( \frac{f^\mathrm{o} - f^\mathrm{e}}{f^\mathrm{o} + 
f^\mathrm{e}} \right)_{\theta = -45^\circ} \right],
\label{eq:circ-pol-wav-pl}
\end{equation}
where $f^\mathrm{o}$ denotes the ordinary and $f^\mathrm{e}$ the
extraordinary beam \citep[see the FORS User Manual for additional
information,][]{fors1+2usermanual04-1}.

Since there were noticeable seeing variations during the observing
run, we applied a correction for the flux loss due to the finite slit
width of 1\arcsec, using the measured FWHM of the object spectrum at
\mbox{5575\,\AA}\ to estimate the effect of seeing and
assuming a Gaussian intensity distribution across the slit.


\begin{figure}[t]
\includegraphics[bb=18 34 564 742,width=8.8cm,clip]{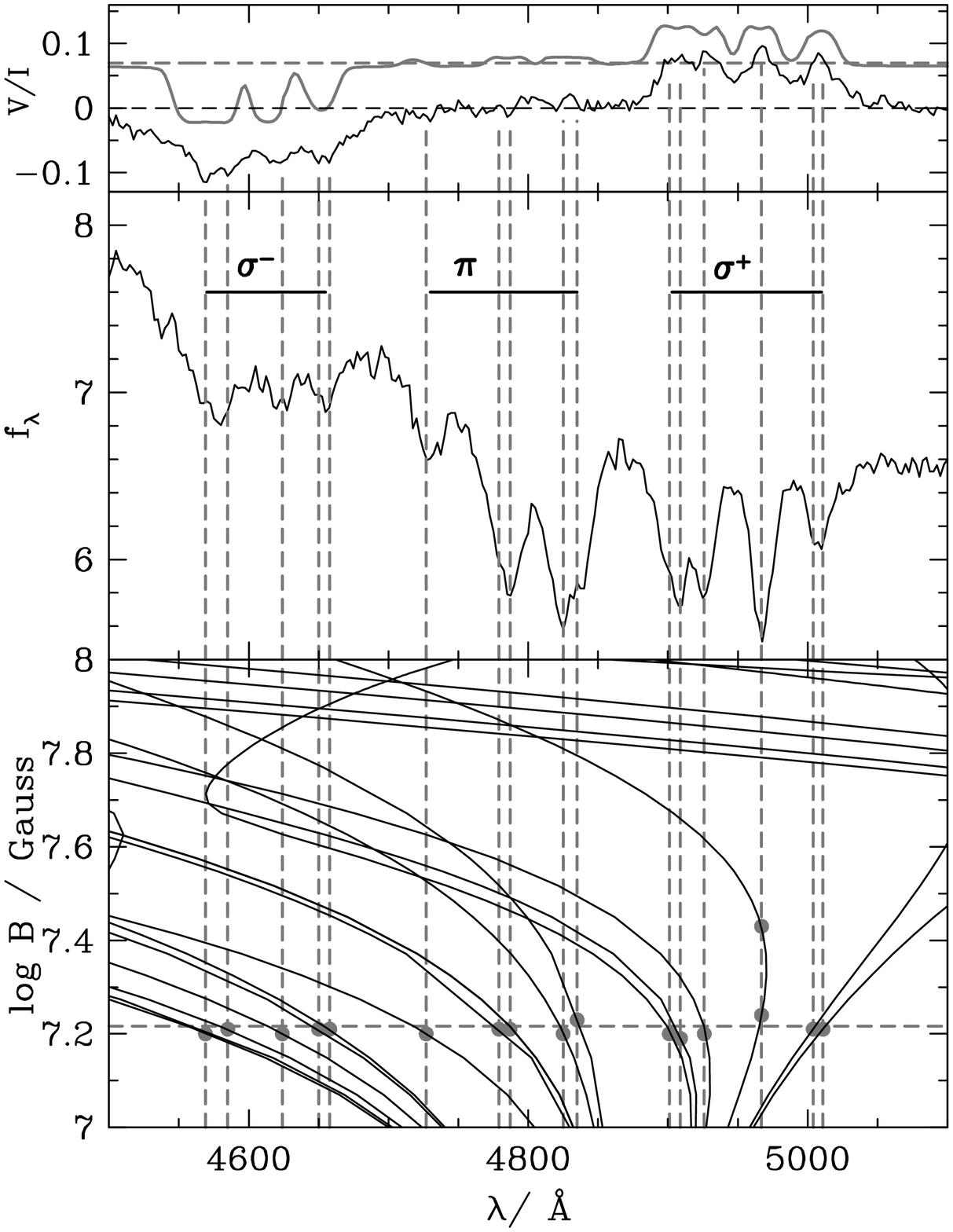}
\caption{Zeeman maximum \mbox{($\phi=0$)} flux $f_\lambda$ (in units
of \mbox{$10^{-16}$ erg cm$^{-2}$ s$^{-1}$ $\AA^{-1}$}) and circular
polarization $V/I$ of \ohe, plotted along with the theoretically
predicted field-dependent transition wavelengths (\lbc) for Balmer
absorption lines.  The top panel also shows a theoretical
circular polarization spectrum for a model atmosphere permeated by a
homogeneous field of \mbox{$B=16$\,MG} (shifted upwards by
0.07 units, with the horizontal dashed grey line denoting the zero
polarization level).  In the bottom panel, filled circles denote
unambiguous identifications of transitions.}
\label{fig:obs-zeeman-maximum}
\end{figure}

\subsection{Data analysis}
\label{sec:data_analysis}

In June~1999, we obtained a sequence of five exposures of \ohe\
covering a time interval of 1.7\,h, terminated by bad weather.  This
run yielded two independent circular polarization spectra
(Eq.~\ref{eq:circ-pol-wav-pl}).  In December~1999, another ``snapshot''
of two additional exposures was secured, yielding another
polarization spectrum.
All flux spectra are shown in Fig.~\ref{fig:he1045-raw-obs} (left
panel).  The temporal change in the five June~1999 spectra is
similar to that seen in the data set of \citet{schmidtetal01-1}.
In our data, the Zeeman features are most
prominent at the beginning of the observations, in the
\citeauthor{schmidtetal01-1} data at the end. 
The features at the beginning and at the end of our run resemble those
at the end and the beginning of the \citeauthor{schmidtetal01-1} run,
respectively, 
i.\,e. the variation of the Zeeman features in
our run is reversed with respect to the \citeauthor{schmidtetal01-1}
data. Our isolated observation in December~1999 also fits into this
pattern.
We estimate that the phase of the strongest Zeeman features
occurs between our spectra~1 and 2 in
Fig.~\ref{fig:he1045-raw-obs}. Spectrum~5 approximately corresponds to
the phase with the weakest Zeeman features.  
The implied
rotational period is \mbox{\prot\ $\simeq 2.7$\,h} if the combined
\citeauthor{schmidtetal01-1} and present observations cover a full rotational 
period. 
While
the substantial variation of the Zeeman features suggests this might
be true, there is unfortunately no proof of such a 
connection of the phase intervals covered in the separate
observations.  
Alternatively, it is possible that the 
\citeauthor{schmidtetal01-1} and our data do not cover a full rotational 
period and
\mbox{\prot\ $> 2.7$\,h}. 
We refer to the former as case~(i) and to the
latter as case~(ii). In case~(ii), there are several possibilities for the
phase intervals covered by our and the \citeauthor{schmidtetal01-1} data 
which we discuss below. 
The large variation in the strength of the Zeeman
features cannot arise in too short a phase interval, however, and we
find that for periods in excess of \mbox{9\,h} an acceptable solution
is no longer
obtained. 
We follow the suggestion of \citeauthor{schmidtetal01-1} of a period in
the \mbox{2--4\,h} range and adopt \mbox{2.7\,h} as the preferred period, 
but report on the consequences of assuming a longer period below.

In preparation of the analysis, we note that the flux
spectra of December~1999 (spectra~6 and 7 in
Fig.~\ref{fig:he1045-raw-obs}) are very similar in shape to
spectrum~5 and provide an additional independent circular
polarization spectrum
which connects in phase to the June~1999 run.
We collect spectra 1/2, 3/4, and 6/7 into three flux
and circular polarization rotational phase bins. For case~(i) with
\mbox{\prot\ $\simeq 2.7$\,h}, these bins are approximately centred
at rotational phases \mbox{$\phi$ = 0.0}, 0.25, and 0.5, where
\mbox{$\phi$ = 0}
refers to the phase of maximum strength of the Zeeman features in our
June~1999 observation. The case~(i) concatenation of our and the
\citeauthor{schmidtetal01-1} data requires that the flux spectrum at
\mbox{$\phi$ = 0.75}
resembles that at \mbox{$\phi$ = 0.25}.
As a representative case~(ii), we
consider twice the rotational period and tentatively assign the three
spectra to \mbox{$\phi$ = 0.0}, 0.125, and 0.25.
Our observations now cover a
phase interval of only \mbox{$\Delta \phi$ = 0.25}.
Furthermore, we
adopt \mbox{\prot\ $\simeq 7.5$\,h} or even \mbox{11.3\,h} and assign the
spectra to \mbox{$\phi$ = 0.0}, 0.09, and 0.18 or 0.0, 0.06, and 0.12 with
\mbox{$\Delta \phi$ = 0.18} or 0.12, respectively.

In Fig.~\ref{fig:he1045-raw-obs} (right panel), we
present the three flux and circular polarization spectra which form
the basis of our tomographic analysis.  We refer to the mean of
spectra 1 and 2 as ``Zeeman maximum'' (\mbox{$\phi = 0.0$}) and to the mean
of spectra 6 and 7 as ``Zeeman minimum'', 
which corresponds to \mbox{$\phi$ = 0.5} for case~(i) and to the smaller
values given above for case~(ii).


\section{Qualitative analysis of the magnetic field geometry}

The rich Zeeman spectra of \ohe\ allow us to obtain insight into the
magnetic field geometry already by the simple means of comparing the
spectra with the expected field-dependent wavelengths of the hydrogen
transitions $\lambda^{\textrm{H}}(B)$, henceforth referred to as \lbc\
\citep{forsteretal84-1, roesneretal84-1, wunneretal85-1}.
Fig.~\ref{fig:obs-zeeman-maximum} shows the wavelength range around
H$\beta$ of the Zeeman maximum spectrum \mbox{($\phi=0.0$)} along with
the \lbc. Several transitions that can be immediately identified are
marked by filled grey circles. This holds for the $\sigma$ and $\pi$
components in the flux spectrum and the $\sigma$ components in the
polarization spectrum, while the circular polarization of the $\pi$
component vanishes, indicating a small viewing angle $\psi$ between
the magnetic field direction and the line of sight.  The distribution
of field strengths is sharply concentrated at \mbox{$\sim$\,16\,MG},
as demonstrated also by the fair agreement between the observed and
the model polarization spectrum shown at the very top of
Fig.~\ref{fig:obs-zeeman-maximum}.  This model spectrum is calculated
for a \emph{single} value \mbox{$B=16$\,MG}, \mbox{$\psi=29$\degr},
and \mbox{\teff\ = 10\,000\,K}. Hence, the field over the visible
hemisphere at this phase is \mbox{$\sim$\,16\,MG} and directed towards
us.

The identification of Zeeman transitions is not as easily
possible in the Zeeman minimum spectrum (\mbox{$\phi=0.5$}), and
we do not show the corresponding attempt of a quick analysis.
Four transitions can definitely be identified, however,
from almost stationary parts of the \lbc\ in H$\alpha$ and H$\beta$
$\sigma^+$, and the corresponding field strengths span a range
from 20 to 60\,MG. This simple analysis proves already that the field
strength over the stellar surface varies by about a factor four,
excluding simple field configurations like a centred or a moderately
offset dipole.


\begin{figure}[t]
\includegraphics[bb=18 34 564 742,width=8.8cm,clip]{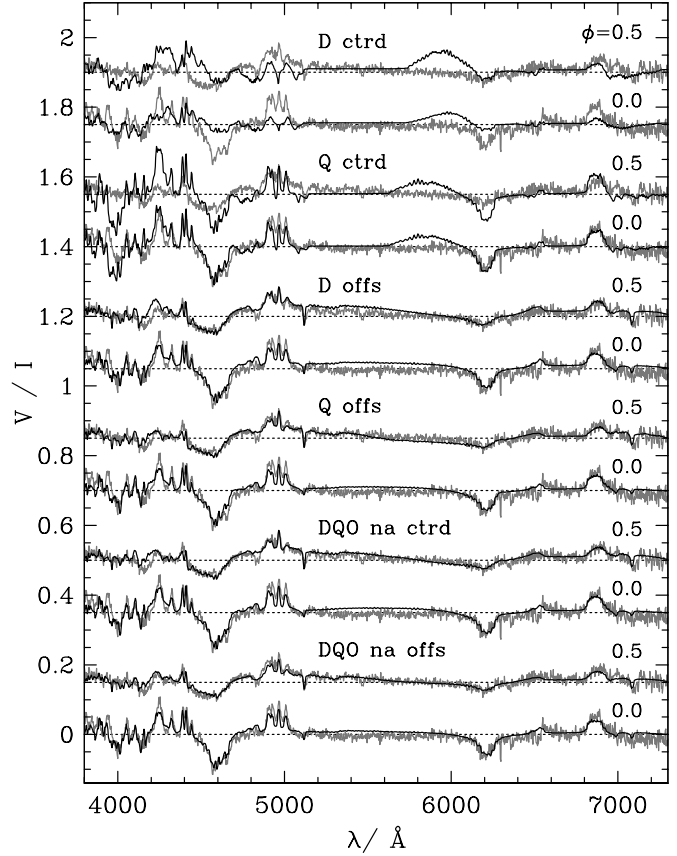} 
\caption{Observed circular polarization spectra for phases
\mbox{$\phi=0.0$} and 0.5 (grey curves) and best-fit synthetic spectra
(black curves) for different parametrizations of the magnetic field
geometry. From top to bottom: centred dipole (D~ctrd); centred
quadrupole (Q~ctrd); off-centred dipole (D~offs); off-centred
quadrupole (Q~offs); centred, non-aligned combination of dipole,
quadrupole, and octupole (DQO~na~ctrd); off-centred, non-aligned
combination of dipole, quadrupole, and octupole (DQO~na~offs).  All
curves except for the bottom one have been shifted vertically by
suitable amounts in $V/I$, with the horizontal dashed lines indicating
the respective levels of zero polarization. 
The quoted phases refer to 
case~(i) with \mbox{\prot\ = 2.7\,h}.}
\label{fig:he1045-result-6ph}
\end{figure}

\begin{table*}[tb]
\caption{Best-fit magnetic parameters for the different
parametrizations of the magnetic field shown in
Fig.~\ref{fig:he1045-result-6ph}. The uncertainties in the last digit
are denoted by the values in brackets. A short description
of the individual models is followed by the formal value of the
reduced $\chi^2$ (see text). All models assume a rotational period of
\mbox{2.7\,h}, except for model (7), which has been computed for 
\mbox{5.4\,h}.}
\label{tab:fit-parameters}
\begin{tabular}{@{\hspace{5mm}}crrrrrrrrrrrr} 
\hline \hline \noalign{\smallskip}

$i$ & \bdip & \thed & \phid & \bqua & \theq & \phiq & 
\boct & \theo & \phio & \xoff & \yoff & \zoff \\

 (\degr) & (MG) & (\degr) & (\degr) & (MG) & (\degr) & (\degr) &
(MG) & (\degr) & (\degr) & (\rwd) & (\rwd) & (\rwd) \\ 

\noalign{\smallskip} \hline
\noalign{\smallskip}


\multicolumn{13}{l}{\textit{(1) D ctrd (centred dipole, \chisqred\ = 121.9)}} \\

 57\,(3) & 
 $-$32\,(1) & 14\,(1) & 30\,(1) &  
 \multicolumn{1}{c}{--} & \multicolumn{1}{c}{--} & \multicolumn{1}{c}{--} &
 \multicolumn{1}{c}{--} & \multicolumn{1}{c}{--} & \multicolumn{1}{c}{--} & 
 \multicolumn{1}{c}{--} & \multicolumn{1}{c}{--} & \multicolumn{1}{c}{--} \\[0.7ex]

 \multicolumn{13}{l}{\textit{(2) Q ctrd (centred quadrupole, \chisqred\ = 104.9)}} \\

 11\,(2) & 
 \multicolumn{1}{c}{--} & \multicolumn{1}{c}{--} & \multicolumn{1}{c}{--} &  
 $-$36\,(1) & 14\,(1) & 18\,(1) &
 \multicolumn{1}{c}{--} & \multicolumn{1}{c}{--} & \multicolumn{1}{c}{--} & 
 \multicolumn{1}{c}{--} & \multicolumn{1}{c}{--} & \multicolumn{1}{c}{--} \\[0.7ex]

 \multicolumn{13}{l}{\textit{(3) D offs (off-centred dipole, \chisqred\ = 29.7)}} \\

 20\,(3) & 
 $-$27\,(3) & 41\,(5) & 29\,(2) &  
 \multicolumn{1}{c}{--} & \multicolumn{1}{c}{--} & \multicolumn{1}{c}{--} &
 \multicolumn{1}{c}{--} & \multicolumn{1}{c}{--} & \multicolumn{1}{c}{--} & 
 0.08\,(1) & 0.02\,(1) &  0.27\,(1) \\[0.7ex]

 \multicolumn{13}{l}{\textit{(4) Q offs (off-centred quadrupole, \chisqred\ = 27.2)}} \\

 18\,(3) & 
 \multicolumn{1}{c}{--} & \multicolumn{1}{c}{--} & \multicolumn{1}{c}{--} &  
 $-$49\,(4) & 20\,(1) & 20\,(2) &
 \multicolumn{1}{c}{--} & \multicolumn{1}{c}{--} & \multicolumn{1}{c}{--} & 
 $-$0.13\,(1) & 0.01\,(1) & 0.06\,(1) \\[0.7ex]

 \multicolumn{13}{l}{\textit{(5) DQO na ctrd (non-aligned, centred combination of dipole, quadrupole, and octupole, \chisqred\ = 26.8)}} \\

 20\,(3) & 
 $-$12\,(2) & 39\,(4) & 17\,(2) &  
 $-$45\,(4) & 36\,(3) & 34\,(6) &
 $-$19\,(1) & 60\,(5) & 27\,(3) & 
 \multicolumn{1}{c}{--} & \multicolumn{1}{c}{--} & \multicolumn{1}{c}{--} \\[0.7ex]

\multicolumn{13}{l}{\textit{(6) DQO na offs (non-aligned, off-centred combination of dipole, quadrupole, and octupole, \chisqred\ = 24.5)}} \\

 17\,(3) & 
 $-$16\,(2) & 71\,(5) & 344\,(2) &  
 $-$36\,(3) & 21\,(3) & 138\,(6) &
 $-$18\,(2) & 70\,(9) & 115\,(7) & 
  0.07\,(1) & $-$0.08\,(1)  & 0.31\,(2) \\

 \multicolumn{13}{l}{\textit{(7) DQO na offs (non-aligned, off-centred combination of dipole, quadrupole, and octupole, \mbox{\prot\ = 5.4\,h}, \chisqred\ = 26.3)}} \\

 32\,(4) & 
 $-$13\,(1) & 31\,(4) & 1\,(1) &  
 $-$37\,(3) & 24\,(3) & 209\,(4) &
 $-$25\,(3) & 69\,(12) & 245\,(10) & 
  $-$0.08\,(1) & 0.07\,(1)  & 0.27\,(2)  \\[0.7ex]

\noalign{\smallskip} \hline

\end{tabular}
\end{table*}


\section{Zeeman tomography of the magnetic field}

Theoretical wavelength-dependent Stokes $I$ and $V$ profiles of
magnetized white dwarf atmospheres can be computed by solving the
radiative transfer equations for given $B$, $\psi$, \teff, \logg, and
the direction cosine $\mu = \cos \vartheta$, where $\vartheta$ denotes
the angle between the normal to the surface and the line of sight.
A synthetic
spectrum for a given magnetic topology can be described by a
superposition of model spectra computed for different parameter
values. 
Our three-dimensional grid of 46\,800 $I$ and $V$ model spectra covers
400 $B$ values (1--400\,MG, in 1\,MG steps), nine $\psi$ values
(equidistant in \mbox{$\cos \psi$}), and 13 temperatures
\mbox{(8000--50\,000\,K)} for fixed \mbox{\logg\ = 8} and \mbox{$\mu =
1$} (Paper I).
This database allows fast computations of synthetic spectra for any
given magnetic field configuration without the need to solve the
radiative transfer equations each time. Limb darkening is accounted
for in an approximate way by the linear interpolation
\begin{equation}
I_{\lambda}(\mu) / I_{\lambda,\mu = 1} = a + b \mu\,.
\label{eq:ld}
\end{equation}
For the sake of simplicity, the temperature- and wavelength-dependencies
of $a$ and $b$ have been neglected (see the discussion in Paper~I). 
Fitting the measured absolute flux distribution of \ohe\ with model
spectra computed using the full radiative transfer method, we find an
effective temperature of \mbox{\teff\ = 10\,000 $\pm$ 1000\,K}.  For
this temperature, model spectra computed as a function of $\mu$
suggest $a=0.53$ and $b=0.47$. In the subsequent tomographic analysis,
these values of \teff, $a$, and $b$ were employed and kept constant.

Our Zeeman tomographic code requires an appropriate parametrization of
the magnetic field, such that for every location $\vec{r}$ on the
stellar surface the magnetic field vector $\vec{B}(\vec{r},\vec{a})$
can be computed depending on a parameter vector $\vec{a} = (a_1,
\dots, a_M)$ of $M$ free parameters describing the field geometry.
Best-fit parameters are determined by minimizing a penalty function as
a measure for the misfit between model and observation. For that
purpose, we employed the C programming language library \texttt{evoC}
by Trint \& Utecht
(1994)\footnote{ftp://biobio.bionik.tu-berlin.de/pub/software/evoC/}
which implements an evolutionary minimization strategy. The penalty
function we used is the classical reduced \chisq\
\begin{equation}
\chi^2_{\mathrm{red}} (\vec{a}) = \frac{1}{N-M} \sum_{j=1}^N 
\frac{(f_j - s_j(\vec{a}))^2} {{\sigma_j}^2}
\label{eq:chisq}
\end{equation}
with the input data pixels $f_j$, the model data pixels $s_j$, and the
standard deviations $\sigma_j$. We used 1321 pixels per phase for the
individual flux and polarization spectra each, yielding $N=7926$
pixels in total (\mbox{$\lambda = 3900$--7200\,\AA}, \mbox{$\Delta
\lambda = 2.5$\,\AA}). All phases have been equally weighted, and
flux and circular polarization have also been given equal weight.
In order to estimate the statistical noise, a Savitzky-Golay filter
with a width of nine pixels (corresponding to 20\,\AA) has been applied
to the observed spectra. Subsequently, the standard deviations
$\sigma_j$ entering  Eq.~(\ref{eq:chisq}) have been computed from the
differences between the filtered and original spectra for wavelength
intervals of \mbox{250\,\AA}.
The standard form of \chisqred\ was used as a suitable relative
goodness-of-fit measure, but the unavoidable systematic differences
between the observed and theoretical spectra prevent that anything
near \mbox{\chisqred\ $\simeq 1$} can be achieved.

In order to avoid that such systematic differences influence the
analysis of the narrower Zeeman structures, we adjusted the model flux
spectra to the observed spectra at a number of wavelengths outside
obvious Zeeman features. This procedure improves the fit of the flux
spectra to the data but does not affect the polarization spectra. The
wavelengths in question are marked by ticks at the top of
Fig.~\ref{fig:he1045-raw-obs} (right panel).  Due to the finite
exposure times, a model spectrum corresponding to a given observed
spectrum should in principle be computed from several model spectra
covering the respective phase interval.  Our code is able to account
for this ``phase smearing'' effect, but the need to compute the
additional spectra slows down the minimization procedure so much that
we decided against this approach. After having obtained best-fit
parameters, we computed spectra including the effect and found no
significant differences. The statistical errors of the best-fit
parameters have been computed using the method described in
\citet{zhangetal86-1}.

Details of the radiative transfer calculations, the synthetisation of
model spectra, the geometry adopted for the description of the
magnetic field configuration, and the fitting strategy are given in
Paper~I.

\subsection{Field parametrization}

In theoretical terms, a magnetic field of very general shape (with the
constraints that it is curl-free and generated only in the stellar
interior) can be described by expanding the scalar magnetic potential
in spherical harmonics depending on the indices $l$ and $m$ for the
degree and order of the expansion \citep{gauss38-1,langel87-1}.  For
each \mbox{$(l,m)$-combination}, two free parameters \glm\ and \hlm\
are assigned, while only one parameter, $g_l^0$, describes the zonal
components with \mbox{$m=0$}.
Thus, the number of free parameters for an expansion up to degree $l$
is \mbox{$l(l+2) = 15$~(24, 35)} for \mbox{$l=3$~(4, 5)}, increasing
rapidly with the maximum degree~$l$. Another property of multipole
expansions is the dependence of the degree $l$ required for an
adequate description on the choice of the reference axis.  Consider,
e.g., a non-aligned superposition of a dipole and a quadrupole, which
can be described exactly by four parameters ($g_1^0$, $g_1^1$,
$h_1^1$, $g_2^0$) \emph{if} the reference axis coincides with the axis
of symmetry of the quadrupole. If the reference axis points in a
different direction, a finite $l$ allows only an approximate
description.

In order to ensure stable and fast convergence of our optimization
scheme, it is necessary to minimize the number of free parameters.  We
adopt, therefore, a hybrid model which implements a superposition of
zonal (\mbox{$m=0$}) harmonics only, disregarding the other tesseral
components with \mbox{$m \neq 0$}.  We allow for arbitrary tilt angles
of the zonal components and also for off-centre shifts.
With this configuration, it is possible to describe fairly complex
geometries with fewer parameters than in the truncated multipole
expansion which includes all tesseral components (see Paper~I for a
detailed description).

\subsection{Results}
\label{sec:results}

\subsubsection{Case~(i): \mbox{\prot\ = 2.7\,h}}

\begin{figure*}[!tH]
\centering
\includegraphics[width=15.9cm,clip]{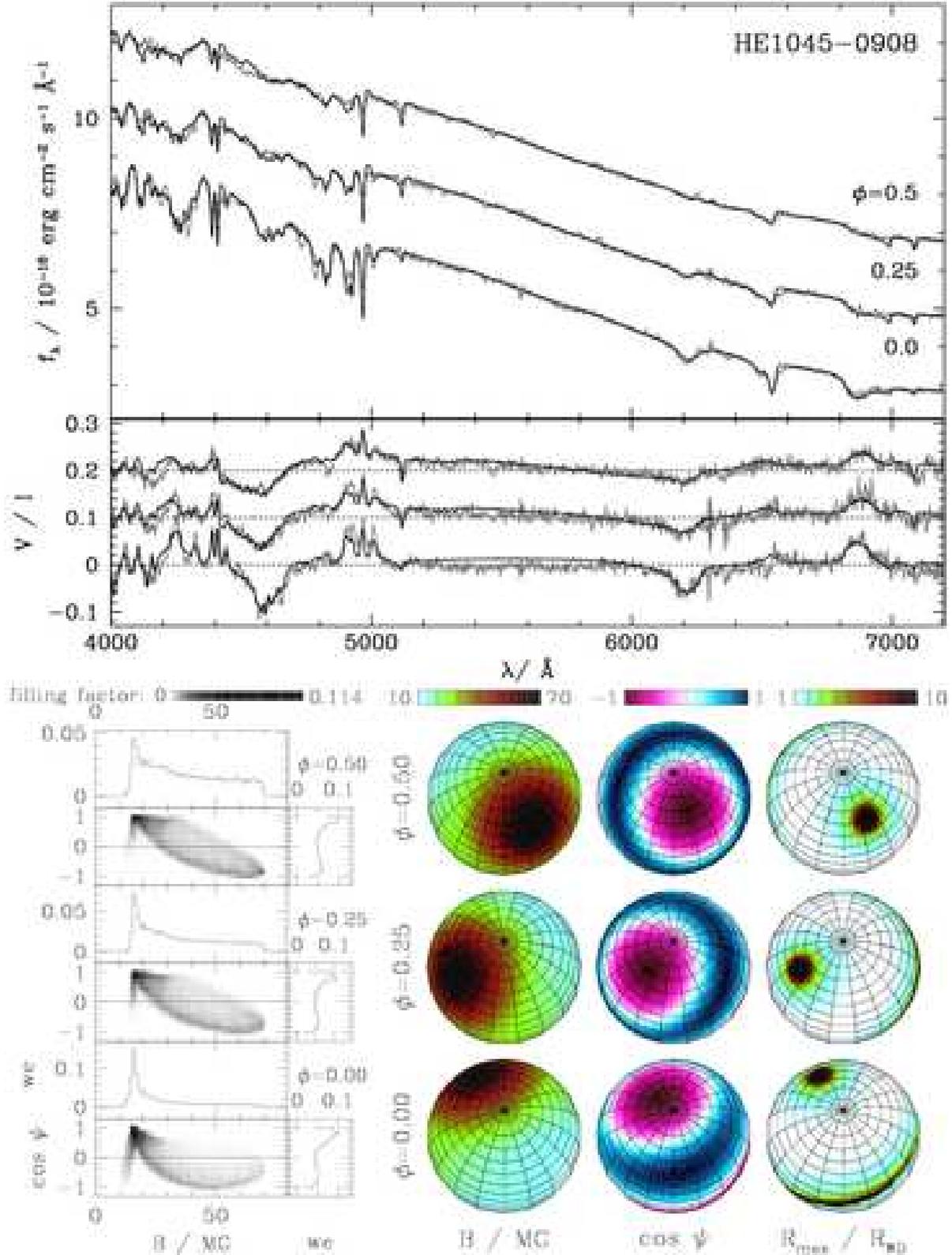} 
\caption{Zeeman tomographic analysis of the magnetic field
configuration of \ohe\ using a centred, non-aligned combination of
dipole, quadrupole, and octupole. \textit{Top:} Observed (grey curves)
and best-fit synthetic spectra (black curves). The uppermost two flux
(circular polarization) spectra have been shifted for clarity by 2 and
4 (0.1 and 0.2) units in $f_\lambda$ ($V/I$). 
The quoted phases refer to 
case~(i) with \mbox{\prot\ = 2.7\,h}. 
\textit{Bottom left:}
\bpd, \textit{bottom right:} absolute value of the surface magnetic
field, cosine of the angle $\psi$ between the magnetic field direction
and the line of sight, and maximum radial distance reached by field
lines in units of the white dwarf radius (see text).}
\label{fig:he1045-result-ctrd}
\end{figure*}

\begin{figure*}[!tH]
\centering
\includegraphics[width=15.9cm,clip]{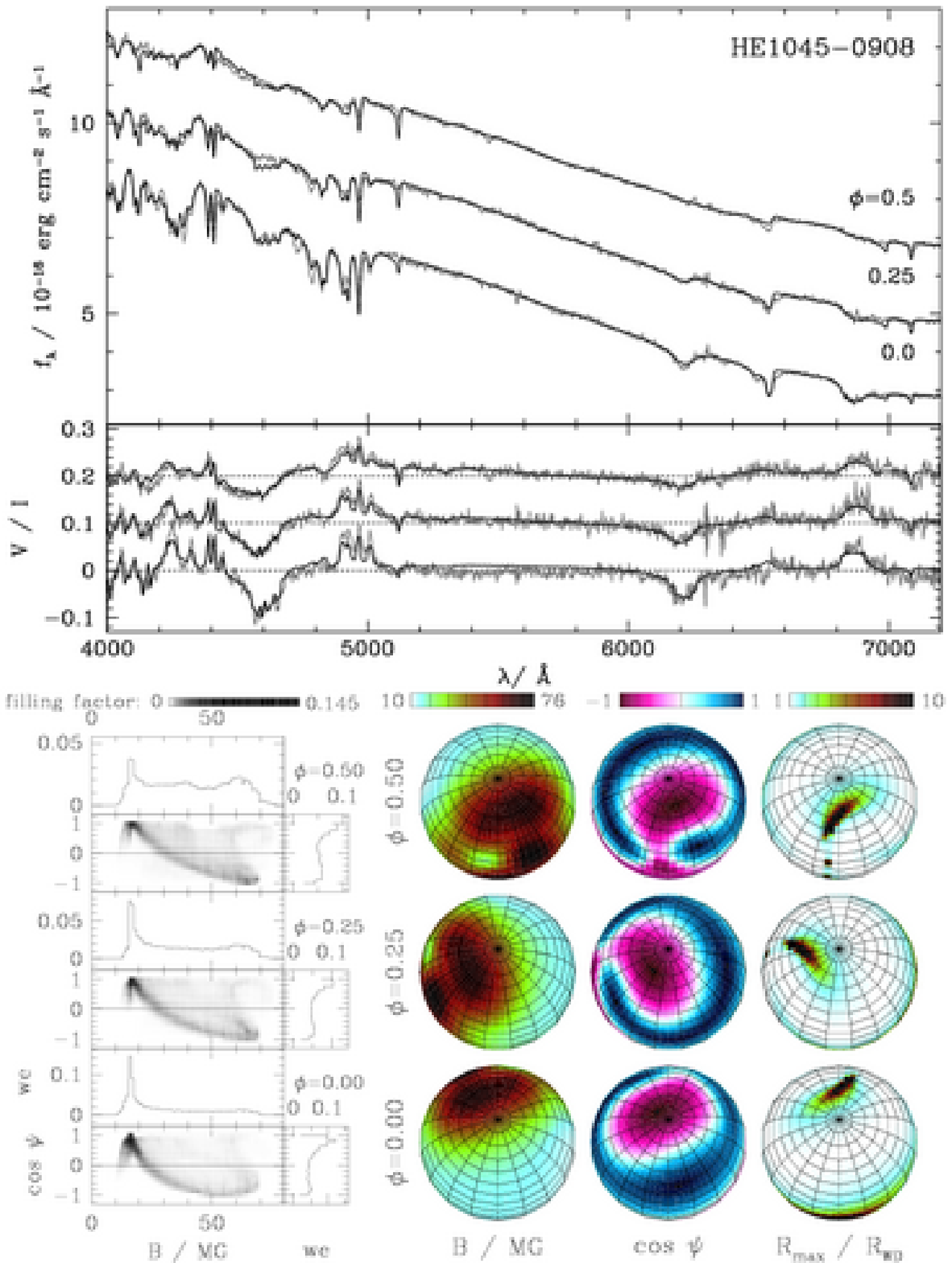} 
\caption{Zeeman tomographic analysis of the magnetic field
configuration of \ohe\ using an off-centred, non-aligned combination
of dipole, quadrupole, and octupole. \textit{Top:} Observed (grey
curves) and best-fit synthetic spectra (black curves). The uppermost
two flux (circular polarization) spectra have been shifted for clarity
by 2 and 4 (0.1 and 0.2) units in $f_\lambda$ ($V/I$). 
The quoted phases refer to case~(i) with \mbox{\prot\ = 2.7\,h}. 
\textit{Bottom left:} \bpd, \textit{bottom right:} absolute value of the surface
magnetic field, cosine of the angle $\psi$ between the magnetic field
direction and the line of sight, and maximum radial distance reached
by field lines in units of the white dwarf radius (see text).}
\label{fig:he1045-result-offs}
\end{figure*}

In an attempt to find the best-fitting field geometry for \ohe,
we considered a sequence of parametrizations with increasing
complexity. In Fig.~\ref{fig:he1045-result-6ph}, we compare the
observed circular polarization spectra at $\phi=0.0$ and
\mbox{$\phi=0.5$} with such a sequence of model spectra.  The best-fit
parameters and the corresponding \chisqred\ values are listed in
Table~\ref{tab:fit-parameters}. For the two simplest configurations
(centred dipole and centred quadrupole) we obtained no satisfactory
fit to the observations. This can be easily explained by the range of
field strengths generated by these configurations, which is too large
for \mbox{$\phi=0.0$} and too small for \mbox{$\phi=0.5$}.
It is interesting to note that the centred dipole, which obviously provides
the least adequate description, is the only configuration that yields an
inclination of \mbox{$i>20\degr$}, whereas for all other configurations an
inclination of \mbox{$i \simeq 10\degr$}--20\degr\ is obtained.

If an appropriate offset from the stellar centre is introduced for the
dipole and quadrupole configurations, the possible range of surface
field strengths increases and an extended region with a nearly
constant field strength of 16\,MG can be generated. Simultaneously,
on the opposite stellar hemisphere a smaller high-field region with a
steeper field gradient is created. As expected, the off-centred
dipole and quadrupole models match the observations better than the
centred configurations, with the quadrupole model fitting better than
the dipole. In general, however, these simple models are unable
to produce adequate fits to all details at all phases simultaneously.

The next steps in complexity of the field configuration are
represented by the superposition of dipole, quadrupole, and octupole
and the introduction of an off-centre shift.  Dipole-quadrupole
combinations were not successful and the inclusion of the octupole is
essential.  

In a first attempt, we allowed the three individual components
to be inclined with respect to each other, but not to be offset from
the centre. The best fit with this field parametrization matches the
observed flux and polarization spectra well for all rotational phases
(Fig.~\ref{fig:he1045-result-ctrd}, top panel).  The frequency
distribution of field strengths extends from 10 to 70\,MG and peaks at
16\,MG. At \mbox{$\phi=0.0$} the distribution of field strengths drops
steeply towards lower \emph{and} higher fields, while at
\mbox{$\phi=0.5$} the peak is less pronounced and the distribution is
much broader, implying that fields up to 70\,MG contribute
significantly to the Zeeman spectra. The \bpd\
(Fig.~\ref{fig:he1045-result-ctrd}, bottom left panel) shows that for
the fields above 30\,MG the sign of $\cos \psi$ is reversed compared
with the most frequent field of 16\,MG. The picture of the field
geometry (Fig.~\ref{fig:he1045-result-ctrd}, bottom right panel) shows
a high-field spot with $B$ up to 70\,MG superimposed on a low-field
background of \mbox{10--20\,MG}. The field geometry on the visible
part of the stellar surface is quadrupole-like with field lines
leading from the high-field pole to an ``equatorial'' band.
The field strengths and orientations of the individual components are
given in line~(5) of Table~\ref{tab:fit-parameters} (see also Eq.~7 in
Paper~I). With \mbox{45 $\pm$ 4\,MG}, the quadrupole is more than
three times as strong as the dipole with \mbox{12 $\pm$ 2\,MG}.  The
three field components are more or less aligned, with the quadrupole
inclined by only 11\degr, and the octupole by 22\degr\ with respect to
the dipole.
The slight inclinations of the individual components with respect to
each other produce the required widening of the high-field spot.
Basically, the field structure is that of an oblique rotator with an
angle of \mbox{$\sim$\,40\degr}\ between field and rotational axes.
The lower right panel, labelled $R_\mathrm{max} / R_\mathrm{WD}$,
indicates the maximum radial distances to which field lines extend in
units of the white dwarf radius. Distances beyond \mbox{10\,\rwd}\
(black) may denote open field lines\footnote{This information may not
be relevant for \ohe\ but is useful in studies of accreting white
dwarfs, because it allows the identification of regions that are
potential foot points of field lines involved in channeled
accretion. Accreting white dwarfs form a part of our programme and
will be dealt with in forthcoming publications.}.

A slightly better fit 
is obtained if we allow for a common offset from the centre for all
three components.
As can be seen from Fig.~\ref{fig:he1045-result-offs} (top panel),
this additional freedom leads, in particular, to improvements in the
model circular polarization which we consider significant: (i) the
steep rise at \mbox{4170--4220\,\AA}\ for \mbox{$\phi=0.0$}; (ii) the
dips at \mbox{4790--4870\,\AA}\ and at \mbox{5300\,\AA}\ for
\mbox{$\phi=0.5$}; and (iii) the continuum polarization in the
\mbox{5200--6000\,\AA}\ range.
The \bpd\ (Fig.~\ref{fig:he1045-result-offs},
bottom left panel) shows an enhanced frequency of field strengths
around 60\,MG for all three phases. For \mbox{$\phi=0.5$} we still see
a pronounced decrease at 70\,MG, but, in contrast to the previous
configuration, there is a small but significant contribution from
fields of \mbox{70--76\,MG} and the same direction as the prevailing
field of 16\,MG.  
Figure~\ref{fig:he1045-result-offs} (bottom right
panel) shows a field geometry that is similar from a global point of
view, but reveals a more complex structure in the high-field region
with two separate areas of opposite field direction. 
This general similarity is produced, however, by an arrangement of the
field components which is entirely different from the previous model
(see also the examples given in Paper~I).
The quadrupole is now inclined by 90\degr\ with respect to the
dipole, and the octupole is not far from orthogonal to both. The shift
is primarily in the direction of the dipole ($z'$-axis), and, hence,
shifts the quadrupole perpendicularly to its axis. 
The field strengths quoted for this model in line~(6) of
Table~\ref{tab:fit-parameters} refer to the unshifted components, and
the final surface values can be calculated with some additional
algebra.
The shape of the region with field lines reaching beyond
\mbox{10\,\rwd}\ (black) has changed from a circular spot to an
arc. Field lines still end in a linearly extended region just visible
at the stellar limb for \mbox{$\phi=0.0$}.

All models in lines~(1) through (6) of Table~\ref{tab:fit-parameters} 
have the property
that the \mbox{$\phi=0.75$} Zeeman spectra resemble those at \mbox{$\phi=0.25$},
as required by the proposed concatenation of the \citeauthor{schmidtetal01-1} 
and our
data. Hence, although the \citeauthor{schmidtetal01-1} data have not been used in
the fit, they are approximately reproduced by the models.

The models of lines~(5) and (6) of Table~\ref{tab:fit-parameters} with 9
and 12 free field parameters, respectively, fit better than the
17-parameter model of the full multipole expansion up to $l=3$
presented in a preliminary report \citep{euchneretal05-1}. 
While this multipole expansion provides a better fit, e.\,g., to the
\mbox{5200--6000\,\AA}\ continuum polarization at \mbox{$\phi=0.0$},
it fails more seriously in other places.
It seems that the choice of the arbitrarily oriented zonal components
is more adequate for the case of \ohe.  Models with about a dozen free
magnetic field parameters represent the present limit of our code at
which a stable convergence to the global minimum in the \chisq\
landscape can be achieved.

The general agreement between the observed and synthetic flux and
polarization spectra has reached a high level which indicates that the
theoretical spectra describe the underlying physics of magnetic
atmospheres more or less correctly by now. The remaining differences
can be traced back to a number of sources. On the theoretical side
these are: (i) uncertainties in the absorption coefficients and
approximations in the treatment of the line broadening, in particular,
the treatment of Stark broadening in magnetic atmospheres; (ii) the
finite resolution of the database, currently limited to \mbox{1\,MG},
which causes small wiggles in the spectra. On the observational side
these are: (iii) remaining problems with the flux calibration, i.e. in
the observationally derived response functions, which we have
attempted to correct for by re-normalizing the observed and model
spectra relative to each other; (iv) small errors in the flat fielding
procedure; and (v) uncertainties in the definition of the standard
deviations $\sigma_j$ of spectral flux and polarization which enter
Eq.~(\ref{eq:chisq}) and determine \chisqred.

\subsubsection{Case~(ii): \mbox{\prot\ $> 2.7$\,h}}

We have repeated the analysis for rotational periods
exceeding the preferred value of \mbox{2.7\,h}, in order to investigate whether
the assumption of a longer period, which implies incomplete phase
coverage, leads to a different field structure.
The somewhat
surprising, but also fortunate result is that none of the
investigated models yields a field structure which deviates
substantially from the one derived above. 

We replace the assumed value 0.50 of the phase interval \mbox{$\Delta \phi$}
covered by our data by 0.25, 0.18, and 0.12, corresponding to
rotational periods of 5.4\,h, 7.5\,h, and 11.3\,h. We consider first
the case of \mbox{\prot\ = 5.4\,h}.
The important finding is that the assumption 
\mbox{$\Delta \phi$ = 0.25} 
does not imply the occurrence of a double wave of full period \mbox{5.4\,h} in the
Zeeman features,
but rather a field structure similar 
to case~(i) seen at the larger inclination of \mbox{$i \simeq 32$\degr}.
At least,
this is true for our models which lack multipole components
higher than the octupole. As an example, we list in
Table~\ref{tab:fit-parameters}, line~(7), the parameters for a
non-aligned, off-centred dipole-quadrupole-octupole combination with an assumed
period of \mbox{5.4\,h}, which should be compared with the model in line~(6)
for our preferred period of \mbox{2.7\,h}. For both models, the 
\citeauthor{schmidtetal01-1} and our data are concatenated at the 
phase of Zeeman maximum, but
in line~(6) the combined data cover a full rotational period, and in
line~(7) only half a period. We conclude that the dominance of the
quadrupole and octupole over the dipole is not affected by the
different choice of the rotational period.

The case~(ii) model with \mbox{$\Delta \phi$ = 0.18} requires an
inclination of \mbox{$i$ = 34\degr},
whereas for \mbox{$\Delta \phi$ = 0.12} (with \mbox{$i$
= 53\degr}) a satisfactory fit could no longer be obtained. This
suggests a maximum value of the rotational period of about \mbox{9\,h}.
It is not surprising that for decreasing \mbox{$\Delta
\phi$} the necessary field variation between Zeeman maximum and
minimum can only be produced by larger inclinations 
which allow for a more rapid variation of the Zeeman features.


\section{Discussion}
\label{sec:discussion}

In this study, we have fitted model Zeeman spectra to high-quality
spectropolarimetric data of \ohe\ using our Zeeman tomography code
(Paper~I),
assuming a \emph{bona fide} rotational period of about
\mbox{2.7\,h}.
We have achieved a good fit which reveals a dominant quadrupole
component with additional dipole and octupole contributions. \ohe\ is
the first white dwarf in which a quadrupole component has
been detected so clearly.
This result is found to be robust against
the assumption of a longer rotational period with
an upper limit at about
\mbox{9\,h}.
In our model, the orientations of the axes of dipole, quadrupole,
and octupole have been treated as free parameters, and it turned out
that this freedom is important in obtaining the best fit. 
This
assumption deviates from a truncated multipole expansion with all 
\mbox{$m \ne 0$} components and is justified by its simplicity and ease of
visualization.
We are confident that we have reached a reliable reconstruction of the
general field structure of \ohe.

The most frequent photospheric field strength and direction is
represented by the maximum in the \bpd\ at 16\,MG and positive
cos\,$\psi$. This is also the field which appears most prominently in
the Zeeman spectra, and a cursory analysis would catalogue this star
as ``having a field strength of \mbox{16\,MG}''. Other sections of the
star display field strengths up to \mbox{$\sim$\,75\,MG}, however,
which are less conspicuous in the observed spectra. Considering the
complete information on the field distribution, we find it difficult
to assign either a ``characteristic field strength'' or a ``polar
field strength'' to \ohe. More appropriate would be quotations like:
(i) the most frequent field is 16\,MG; (ii) the mean field over the
visible surface averaged by the surface area is 34\,MG; and (iii) the
range of field strengths is \mbox{10--75\,MG}. Our general experience
is, however, that a quotation of type (iii) is model-dependent because
models for some stars studied by us imply a high-field extension in
the \bpd, which covers only a small area near the limb of the visible
surface and has little statistical significance.

The derived field structure of \ohe\ is primarily defined by the \bpd\
rather than by the strengths and angles of the individual
components. As pointed out in Paper I, different parameter
combinations of the individual components could lead to a similar
\bpd. This is why \citet{donatietal94-1} refrained from specifying
multipole components and suggested to directly optimize the \bpd. The
approach of \citeauthor{donatietal94-1} does not guarantee, however,
that the derived \bpd\ corresponds to a physically possible
field. This potential trap is avoided in our approach, which has the
additional advantage that we can specify the contributions of
individual multipole components.
Furthermore, since we have gradually increased the level of complexity
of our field parametrization starting from the elementary case of a
centred dipole, we can be sure to have found the \emph{simplest}
configuration compatible with the observations.  We cannot exclude
additional small scale structure of the surface field, but suggest that such a
structure cannot dominante \ohe\ because it would destroy
the remarkably high degree of circular polarization of up to
\mbox{$\sim$\,10\,\%}.

Due to the small inclination (\mbox{$i$ = 17\degr}) found for the best-fitting
model geometry, a fraction of \mbox{35\,\%} of the stellar surface is
permanently hidden from view. In Paper~I we have shown that this lack of
information does not affect the accuracy of the derived field structure 
on the visible part of the surface. 
The field structures predicted by all our models
for the hidden part of the surface
are reasonable and ``well-behaved'', 
i.\,e.\ there are no extreme field values or gradients.
For instance, for the case~(i) model of Fig.~\ref{fig:he1045-result-offs}
and line~(6) of
Table~\ref{tab:fit-parameters}, 
the range of field strenghts encountered on the visible fraction
of the surface is \mbox{10--76\,MG}, while for the whole star it is
\mbox{9--76\,MG}.
Our simulations had shown that reconstruction
artefacts can arise if the field parametrization 
involves more free parameters than needed 
to describe the field structure adequately (cf.\ Fig.~10
in Paper~I). 
This does not happen in our stepwise approach with a limited number
of parameters. 

It goes without saying that the final determination of the field structure
of \ohe\ would greatly benefit from a measurement of its rotational period
and full phase coverage of the Zeeman spectropolarimetry.

With \mbox{\teff\ $\simeq 10^4$\,K}, \ohe\ is \mbox{$\sim$\,0.5\,Gyr}
old, less than the Ohmic decay times of the detected multipole
components \citep{wendelletal87-1, cumming02-1}. Hence, the strong
quadrupole component could be a remnant from the main-sequence (or
pre-main-sequence) evolution of the progenitor star. Alternatively, it
could be produced by field evolution in the white dwarf stage as
suggested by \citet{muslimovetal95-1}. 
Now that tomographic methods have reached a high accuracy thanks to
advanced spectropolarimetric instruments at 8-m class telescopes, it
would be interesting to follow up this question by a detailed field
analysis of white dwarfs of different ages, supplemented by more
extensive theoretical calculations of the evolution of white dwarf
magnetic fields over their cooling times.


\begin{acknowledgements}
The referee G.~Mathys provided helpful and valuable
comments and suggested, in particular, to investigate rotational
periods larger than \mbox{2.7\,h}.  
This work was supported in part by BMBF/DLR grant
\mbox{50\,OR\,9903\,6}.  BTG was supported by a PPARC Advanced
Fellowship.
\end{acknowledgements}

\bibliographystyle{aa}

\end{document}